\documentclass[12pt]{iopart}

\usepackage{graphicx}								

\usepackage{harvard}
\usepackage{bm}
\begin{document}

\title[]{Photo-ionization of polarized lithium atoms out of an all-optical atom trap: A complete experiment}
\author{F.~Thini, K.L.~Romans, B.P.~Acharya, A.H.N.C.~de Silva, K.~Compton, K.~Foster, C.~Rischbieter, C.~Russ, S.~Sharma, S.~Dubey, D.~Fischer}

\address{Dept.~of Physics and LAMOR, Missouri University of Science \& Technology, Rolla, MO 65409, United States of America}
\ead{fischerda@mst.edu}
\vspace{10pt}
\begin{indented}
\item[]November 2019
\end{indented}

\begin{abstract}
An all-optical, near-resonant laser atom trap is used to prepare an electronically excited and polarized gas target at mK-temperature for complete photo-ionization studies. As a proof-of-principal experiment, lithium atoms in the 2$^2$P$_{3/2}$($m_l$=+1) state are ionized by a 266\,nm laser source, and emitted electrons and Li$^+$ ions are momentum analyzed in a COLTRIMS spectrometer. The excellent resolution achieved in the present experiment allows not only to extract the relative phase and amplitude of all partial waves contributing to the final state, it also enables to characterize the experiment regarding target and spectrometer properties. Photo-electron angular distributions are measured for five different laser polarizations and described in a one-electron approximation with excellent agreement.  
\end{abstract}

%
%
%
%
%

\section{Introduction}

The photoelectric effect is among the most fundamental, most intensely studied, and at the same time best understood processes in atomic and molecular physics. In particular, the investigation of photo-electron angular distributions (PAD) proved to provide useful information \cite{Manson82,Reid03} because the anisotropy in the electron emission direction is very sensitive to the details of the few-body dynamics and to correlations between the electrons. It is straight-forward to show that PADs for single ionization of initially unpolarized or randomly oriented atomic or molecular targets can generally be expressed by a second order partial wave expansion with only a single system-specific parameter, the so-called anisotropy parameter $\beta_2$ \cite{Cooper68}. Therefore, many early theoretical and experimental studies focused on the investigation of this parameter which contributed tremendously to our understanding of few-body dynamics and light-matter interaction in atomic and molecular systems.

For polarized targets the photo-electron emission profile is characterized by more than a single parameter, thereby providing additional information \cite{Klar82}. In such experiments, the relative amplitudes and phases of several contributing partial waves (specifically for $\Delta l=+1$ and $\Delta l=-1$) can be extracted from the electron angular distributions which is generally not possible for unpolarized targets. While these {\it complete} experiments were pioneered already in the 1990s (e.~g.\  \cite{Pahler92,Becker98,Godehusen98}), new experimental possibilities such as advanced velocity map imaging (VMI) electron spectrometers and the availability of intense lasers and VUV sources further boosted this research area (e.~g.\ \cite{Zhu09,OKeeffe10,Meyer11}) and give new and exciting insights into attosecond electronic dynamics and the details of electronic wave functions \cite{Villeneuve17,Carpeggiani19}.

Here we report on a new experimental method based on a MOTReMi (magneto-optical trap - reaction microscope) setup \cite{Hubele15} equipped with an all-optical atom trap (AOT) \cite{Sharma18}. In this experiment, lithium atoms are trapped, cooled, excited, and polarized with near-resonant laser light. After the ionization of the atoms due to the interaction with single photons or intense laser fields the target fragments, i.e.\ the emitted electrons and recoiling target ions, are momentum analyzed in coincidence using the COLTRIMS technique  (cold target recoil ion momentum spectroscopy) \cite{Doerner00,Ullrich03}. Compared to many earlier COLTRIMS experiments, the present approach is particularly appealing in several respects: First, the ionization of alkali atoms is studied, i.e.\ the binding energy of the active valence electron is among the lowest occurring in nature. That can simplify experimental procedures significantly, because electronic transitions are induced already by optical frequencies instead of VUV or XUV radiation from synchrotrons, free electron lasers, or high-harmonic generation that are required to trigger similar reactions in noble gas atoms. Second, in the AOT optical pumping is employed providing not only an extremely cold target gas cloud, but also a very high degree of target polarization.



As a proof-of-principal experiment, we studied the valence ionization of excited and polarized lithium atoms from the 2p($m_l=1$) state by the absorption of single photons from a pulsed ultra-violet laser source for various polarizations. From theoretical perspective, the description of this photo-ionization processes is particularly simple and straight-forward. It can be closely approximated by electric dipole transitions in an effective one-electron picture neglecting electronic correlation effects. Moreover, due to the small size of lithium, relativistic effects are negligibly small. Therefore, the present data do not challenge theoretical  state-of-the-art models. However, the measured PADs are very sensitive to potential experimental artifacts and systematic irregularities. This makes this experiment an ideal test of the overall quality of the measurement, as well as of the fidelity of the target preparation. Therefore, the present data provide very useful information for future experiments where other processes, e.g. multi-photon ionization on polarized excited and ground-state atoms, inner-shell ionization, or lithium-molecule fragmentation, will be investigated.


As will be shown below, the present measured PADs can be described as a superposition of four partial waves with only two system-specific parameters that either require numerical analysis or can be fitted from the experimental data. The large variety of laser polarizations and the excellent resolution obtained in the present study gives a direct and intuitive picture of the interference of the involved partial waves and, therefore, into the photon-atom interaction. As such, the measured data represent a illustrative 'text-book' example of the fundamentally important atomic photoelectric effect.

\section{Theory}
The  atomic photo-electric effect is a fundamental example of quantum-mechanical scattering theory and the basic approach is described in many atomic physics or quantum mechanics textbooks. Generally, the angular-differential photo-ionization cross section is calculated from the transition matrix element $ D^{if}$ \cite{Bethe77}  with
\begin{eqnarray}
\frac{{\rm d}\sigma}{{\rm d}\Omega _k} \propto \left| D^{if}(\theta_k,\phi_k)\right| ^2 .
\label{eq:crosssection}
\end{eqnarray}
Here $k$ denotes the photo-electron momentum. In the length form and using the dipole approximation, the transition matrix element can be written as \cite{Starace06}
\begin{eqnarray}
D^{if}= \left< \Psi_f \right| \bm{\hat{\epsilon}} \cdot \bm{r} \left| \Psi_i \right>
\label{eq:dme}
\end{eqnarray}
with $\bm{\hat{\epsilon}}$ being the unit polarization vector of the ionizing light field. In the present experiment, the target atom is in an excited state populating a single magnetic sub-level. Describing the atom in a one-electron picture, the expression for the initial and final state wave function $\Psi_i$ and $\Psi_f$ can be approximated by 
\begin{eqnarray}
\psi_{nlm}(\bm{r}) = R_{nl}(r)Y_{l}^m(\bm{\hat{r}})\label{eq:initialstate}\\
\psi_{k}(\bm{r}) = \sum_{l',m'} R_{kl'}(r)Y_{l'}^{m'}(\bm{\hat{r}}) Y_{l'}^{m'*}(\bm{\hat{k}})
\label{eq:finalstate}
\end{eqnarray}
with $R_{nl}$ and $R_{kl}$ being the radial wave functions for the initial bound and final continuum state, respectively.

In order to conveniently evaluate the dipole matrix element in Eq.~\ref{eq:dme}, the dipole operator can be expressed in spherical harmonics, too:
\begin{eqnarray}
\bm{\hat{\epsilon}} \cdot \bm{r} &=&\epsilon_x r \sin \theta \cos \phi + \epsilon_y r \sin \theta \sin \phi + \epsilon_z r \cos \theta  \nonumber
\\ &=& r \sqrt{\frac{4\pi}{3}}\left(\frac{\epsilon_x+i\epsilon_y}{\sqrt{2}}  Y_1^{-1}(\bm{\hat{r}}) + \epsilon_z Y_1^0(\bm{\hat{r}}) + \frac{-\epsilon_x+i\epsilon_y}{\sqrt{2}}  Y_1^{1}(\bm{\hat{r}})\right) \nonumber\\
 &=&r \sqrt{\frac{4\pi}{3}}\left(\epsilon_-  Y_1^{-1}(\bm{\hat{r}}) + \epsilon_z Y_1^0(\bm{\hat{r}}) + \epsilon_+  Y_1^{1}(\bm{\hat{r}})\right) .
\label{eq:dipoleoperator}
\end{eqnarray}
Here, $\epsilon_+$ and $\epsilon_-$ describe the complex relative electric field strength of right-handed (RHC) and left-handed circularly (LHC) polarized light, respectively, propagating in $z$-direction, while $\epsilon_z$ corresponds to linear polarization along the $z$-axis. These three polarizations represent a complete basis set, \emph{i.e.}\ any other polarization can be expressed as a superposition of these three.

Inserting Eqs.~\ref{eq:initialstate} - \ref{eq:dipoleoperator} into equation~\ref{eq:dme} gives for the dipole matrix element 
\begin{eqnarray}
&D^{if}(\bm{\hat{k}})\propto & \sum_{l'=0}^\infty \int{\rm d}r\ r^3\  R_{nl}(r)  R_{kl'}^*(r)\cdot\nonumber\\
&&\int {\rm d}\Omega_r \sum_{m'=-l'}^{l'} Y_l^{m}\left(\epsilon_-  Y_1^{-1} + \epsilon_z Y_1^0 + \epsilon_+  Y_1^{1}\right)Y_{l'}^{m'*}\cdot  Y_{l'}^{m'}(\bm{\hat{k}}).
\label{eq:dipolematrix2}
\end{eqnarray}
For brevity, the dependence of the spherical harmonics on the electron coordinate is not explicitly stated in the equation.

\begin{figure}
\centering
\includegraphics[width=0.6\linewidth]{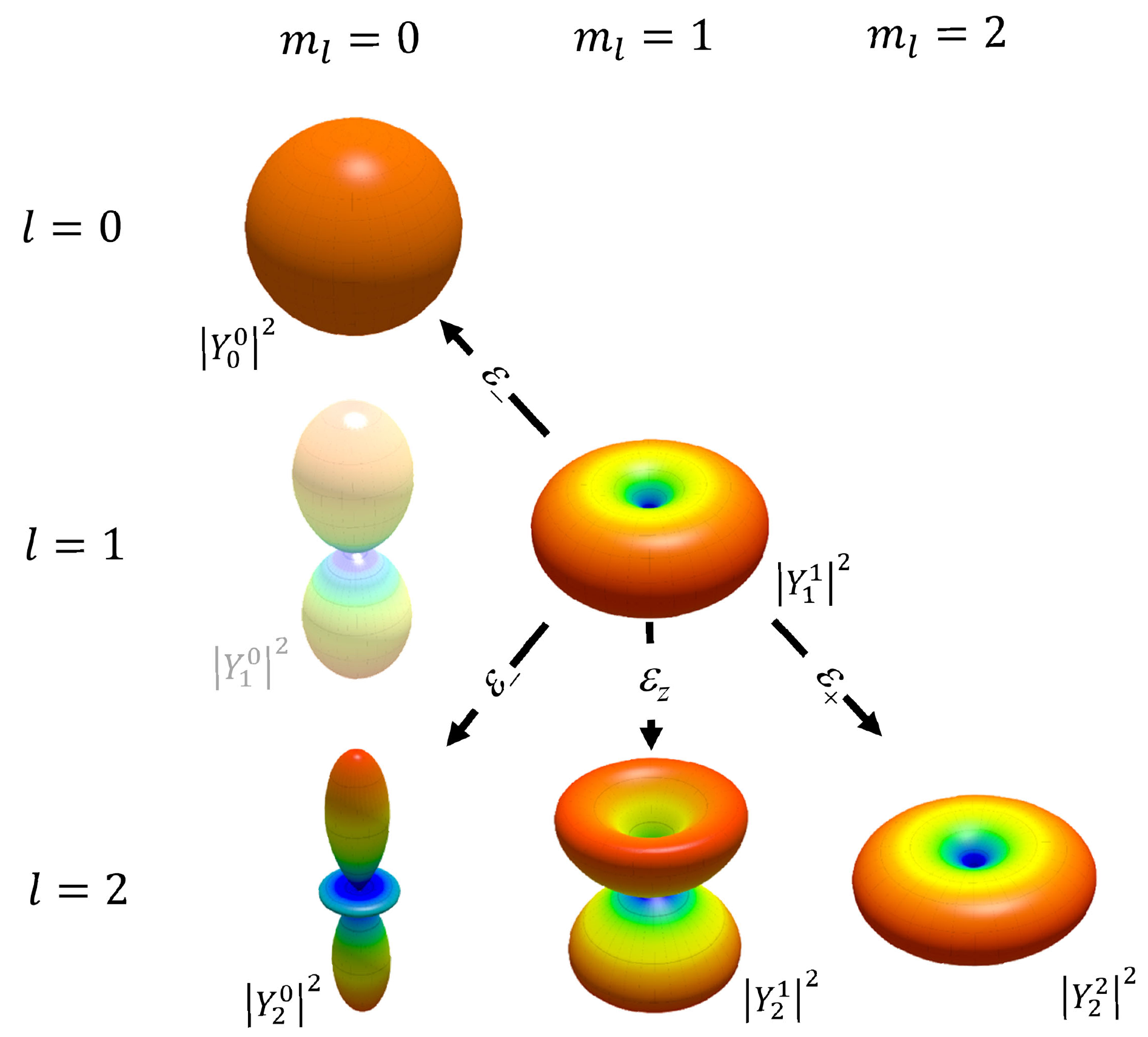}
\caption{Absolute square of spherical harmonics with $l\leq 2$. The arrows indicate the dipole-allowed transitions from the initial state ($Y_1^1$) that are induced by the electric field components $\epsilon_-$, $\epsilon_z$, and $\epsilon_+$. \label{fig:SphHarm}}
\end{figure}

The integration of the angular part of the matrix element is trivial and it yields the well-known dipole-selection rules which are $\Delta l=l-l'=\pm 1$ and $\Delta m=m-m'=-1,0,$ or $+1$ for the first ($\epsilon_-$), second ($\epsilon_z$), or third ($\epsilon_+$) term of the integral, respectively. In the present experiment, the lithium atoms are initially in the 1s$^2$2p($m_l=1$) electronic configuration (see Fig.~\ref{fig:SphHarm}). In this situation, Eqs.~\ref{eq:crosssection} and \ref{eq:dipolematrix2} give a PAD of
\begin{eqnarray}\label{eq:PAD}
&\frac{{\rm d}\sigma}{{\rm d}\Omega _k} \propto& \left| \frac{-\epsilon_- }{2\sqrt{\pi}} A_{02} {\rm e}^{{\rm i} \alpha_{02}} Y_{0}^{0}(\bm{\hat{k}}) 
+\right. \nonumber\\
&& \frac{\epsilon_- }{2\sqrt{5\pi}} Y_{2}^{0}(\bm{\hat{k}})
+\left.  \frac{\sqrt{3}\epsilon_z }{2\sqrt{5\pi}} Y_{2}^{1}(\bm{\hat{k}}) +\frac{\sqrt{3}\epsilon_+ }{\sqrt{10 \pi}} Y_{2}^{2}(\bm{\hat{k}})    \right| ^2 .
\end{eqnarray}

The only parameters in the above expression that depend on the target species and photon energy are the relative amplitude $A_{02}$ and phase $\alpha_{02}$ of the radial part of the matrix element (Eq.~\ref{eq:dipolematrix2}) with $l'=0$ and 2 which are given by
\begin{eqnarray}
A_{02} {\rm e}^{{\rm i} \alpha_{02}}=\frac{ \int{\rm d}r\ r^3\  R_{21}(r)  R_{k0}^*(r)}{ \int{\rm d}r\ r^3\  R_{21}(r)  R_{k2}^*(r)} .
\label{eq:relMat}
\end{eqnarray}
In general for non-Coulombic potentials, the bound and continuum state wave functions and the corresponding radial matrix elements cannot be expressed analytically. In the present study, we solved the Schr{\"o}dinger equation numerically using the Li model potential from \cite{Marinescu1994} in order to calculate the relative amplitude and phase $A_{02}$ and $\alpha_{02}$.



\section{Experiment}

The measurement was performed with a MOTReMi (magneto-optical trap - reaction microscope) setup. While many of the features of the setup have been reported earlier \cite{Hubele15,Sharma18}, some of the details of the target preparation, the momentum spectrometer, and the experimental geometry that are relevant for the present study are discussed in the following.

The $^6$Li atoms are captured and cooled in an all-optical laser atom trap (AOT) \cite{Sharma18} in the crossing region of three pairs of counter-propagating, mutually perpendicular, continuous-wave laser beams with a frequency slightly red-detuned with respect to the 1s$^2$2s $^2$S$_{1/2}$ -- 1s$^2$2p $^2$P$_{3/2}$ cooling transition. The present all-optical trap configuration is similar to magneto-optical traps. However, the AOT does not require a quadrupole magnetic field in the trap region. Instead it can be operated in a weak homogeneous magnetic field (here $B=5.25$\,Gauss). This makes the AOT the ideal target to be used with COLTRIMS which generally requires well-defined homogeneous magnetic fields in the spectrometer region in order to reconstruct electron and recoil ion momentum vectors from their recorded time-of-flights and positions on the particle detectors.

Another important feature of the AOT is that the target atoms are inherently polarized. The high degree of atomic polarization is a result of optical-pumping, because transitions with $\Delta m=+1$ are favored in the field of the cooling lasers. This is not only caused by the choice of laser polarizations, but also by the splitting of the Zeeman levels in the homogeneous field along with the red-detuning of the incoming radiation. A magnetic population distribution of the excited 1s$^2$2p $^2$P$_{3/2}$ state of roughly 0.93, 0.05, and 0.02 for $m_l=+1$, 0, and -1, respectively, was reported earlier (for detail see \cite{Sharma18}). These numbers were obtained by fluorescence spectroscopy measuring the polarization of the re-emitted light. It is worth noting that, due to the optical pumping, electron and nuclear spins are polarized, too, and any influence of spin-orbit couplings on the $m_l$-distribution can largely be neglected in the present system.

The target atoms are photo-ionized with a passively Q-switched Nd:YAG laser operating on the 4$^{th}$ harmonic ($\lambda=266$\,nm) with a pulse duration of about 500\,ps, a repetition rate of 6\,kHz, and an average power of about 1\,mW. In the continuous field of the cooling lasers, about 25\,\% of the target atoms are in the excited  1s$^2$2p $^2$P$_{3/2}$ ($m_l=1$) state.  The photon energy of the ionizing ultra-violet laser (4.66\,eV) suffices to ionize these excited lithium atoms with an excess energy of about 1.1\,eV, while atoms in the ground state remain unaffected due to their higher ionization potential of 5.39\,eV.

The emitted electrons and the recoiling Li$^+$ ions are detected in coincidence using COLTRIMS. In the case of photo-ionization, the electron and ion momentum measurements provide redundant information due to momentum conservation. Therefore, the spectrometer field configuration was chosen for optimal electron momentum resolution. This was achieved with an homogeneous electric field in the entire spectrometer region without employing the commonly used field-free drift region. In this configuration, fringe field effects are minimized and the momentum components can easily be calculated \cite{Fischer19}. The recoil ions were detected with rather moderate momentum resolution (between 0.05 and 0.1\,a.u.), but their detection allowed to reduce background in the electron spectra by applying a coincidence condition of the two fragments. It should be noted that the homogeneous magnetic field results in spiral trajectories of the electrons, and the initial momentum is not unambiguously computable for an electron that undergoes an integer number of cyclotron cycles at the time it hits the detector \cite{Moshammer2003}. Therefore, each measurement was performed for two different electric extraction fields (0.326\,V/cm and 0.277\,V/cm). By combining the data sets for the two field settings, a high electron momentum resolution was achieved throughout the entire final momentum space.

In the discussion below, the following coordinate system is chosen: The $z$-axis is parallel to the magnetic field direction such that the polarization  of the excited target initial 2p-state corresponds to $m_l=+1$. The $x$ and $y$-axes are perpendicular to the $z$-direction with the former being horizontal and the latter vertical in the laboratory frame. Photo-ionization was measured for five different settings of the ultra-violet laser's propagation direction and polarization. Two measurements were performed with right and left circularly polarized light with the laser beam propagating in the $xz$-plane with a small angle of 10$^\circ$ with respect to the $z$-axis. Here, the photo-ionization is dominated by transitions with $\Delta m_l=+1$ and $-1$, respectively. Three data sets were obtained with the laser beam laying in the $xy$-plane and the beam pointing diagonally upwards at an angle off 18$^\circ$ to the $y$-axis. The three polarizations were chosen at 0$^\circ$, 90$^\circ$, and 45$^\circ$ with respect to the $z$-axis. The first polarization corresponds to $\Delta m_l=0$ transitions, the second to a mixture of $\Delta m_l=+1$ and $-1$, and the third polarization results in a superposition of all three possible magnetic quantum numbers.


\section{Results and discussion}

\begin{figure}
\centering
\includegraphics[width=\linewidth]{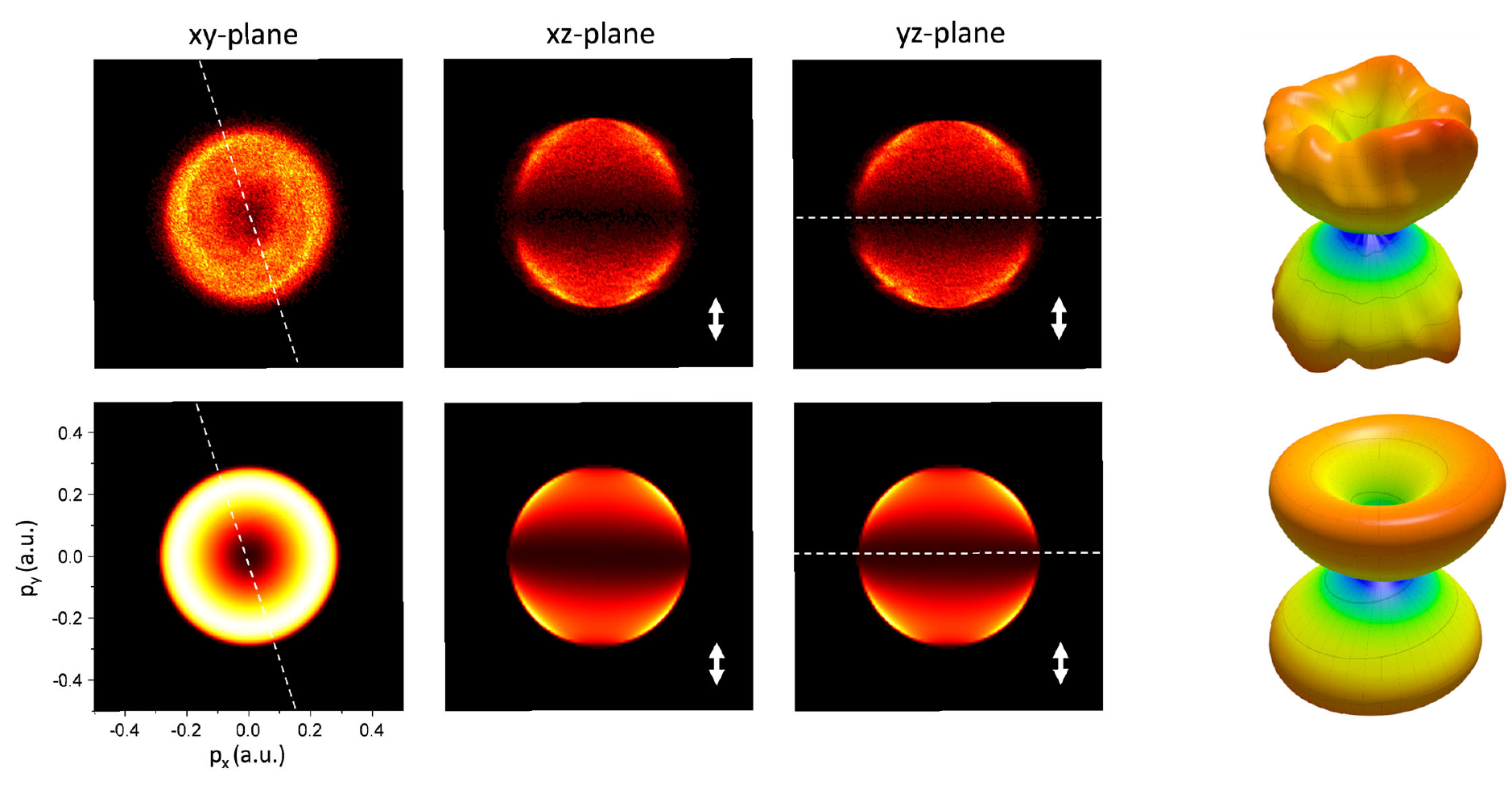}
\caption{Experimental (top row) and theoretical (bottom row) cross sections for the photo-ionization of the Li 2p ($m_l=+1$) valence electron for the photon polarization in the $z$-direction. The left three columns show the electron momentum distribution projected on the $xy$, $xz$, and $yz$-planes. For reference, the momentum scale is shown in the bottom left graph, but omitted in all other graphs. The dashed lines and double arrows indicate the (projected) laser propagation and polarization directions, respectively. The right column represents the electron angular distribution in 3-dimensional polar plots. \label{fig:z}}
\end{figure}

The angular distributions of the photo-electrons emitted after ionization from the 2p ($m_l=+1$) state have been measured for five different incoming laser beam polarizations. As it can be seen from Eq.~\ref{eq:PAD}, the theoretical description of the PADs gets particularly simple, if only one of the three components of the laser electric field vector $\bm{\hat{\epsilon}}$ ($\epsilon_z$, $\epsilon_+$, and $\epsilon_-$) is non-zero, because in this case only one (or two, in the case of $\epsilon_-\neq 0$) partial waves contribute to the final state. Therefore, these laser polarizations are ideally suited to investigate possible experimental artifacts and irregularities. In the first measurement, the laser electric field vector was aimed to be approximately parallel to the $z$-axis (i.e. $\epsilon_+=\epsilon_-=0$). In this case, the change of the magnetic quantum number has to be $\Delta m=0$. Due to the additional requirement of  $\Delta l=\pm 1$ for electric dipole transitions, the final state angular distribution is described by the absolute square of the spherical harmonics $Y_{2}^{1}$. Experimental (top row) and theoretical (bottom row) spectra for this situation are shown in figure \ref{fig:z}. Due to energy conservation, all momenta have to be on a sphere with a radius of about 0.286\,a.u.\ (corresponding to an electron kinetic  energy of 1.1\,eV) in three-dimensional momentum space. The projection of the momentum distribution on the $xy$, $xz$, and $yz$-plane are shown in the left three columns of the figure. In the right column, the PADs are shown as 3-dimensional polar plots.

It should be noted that it is not easily possible to align the polarization of the ionizing laser beam exactly along the $z$-axis in the present epxeriment. On its way into the vacuum chamber the polarization can be altered due to a slight unknown birefringence of the fused-silica viewports \cite{Xiao18}. While this birefringence is largely insignificant at optical wavelengths, it cannot be neglected for ultra-violet radiation. This results in a small systematic uncertainty which is accounted for by slightly modifying the theoretical electric field vector $\bm{\hat{\epsilon}}$ from the values that were measured outside the vacuum chamber in order to achieve the best fit of the calculated PAD to the measured spectra. For the calculation in figure~\ref{fig:z}, we considered the electric field having $\epsilon_+$ and $\epsilon _-$ contributions of about 5\,\%.

As can be seen from the figure, excellent agreement of experiment and calculation is achieved. The experimental electron momentum uncertainty is less than 0.005\,a.u.\ (FWHM) in the $z$-direction and between 0.01 and 0.02\,a.u.\ in the $x$ and $y$-directions which corresponds to an improvement by a factor of 2 to 10 as compared to earlier measurements with the same spectrometer using a switched magneto-optically trapped (MOT) target \cite{Hubele15}. This excellent resolution enables to resolve detailed features of the final state momentum distribution. Some small irregularities for certain $z$-momenta can be observed in the $xz$ and $yz$-plane. Those are artifacts due to the cyclotron motion of the electrons in the magnetic field (e.g.\ \cite{Moshammer2003}) but are much weaker in the present measurement than in most earlier COLTRIMS experiments, because, as discussed in the previous chapter, the experimental spectra were obtained by combining two data sets with different spectrometer field settings.

\begin{figure}
\centering
\includegraphics[width=0.5\linewidth]{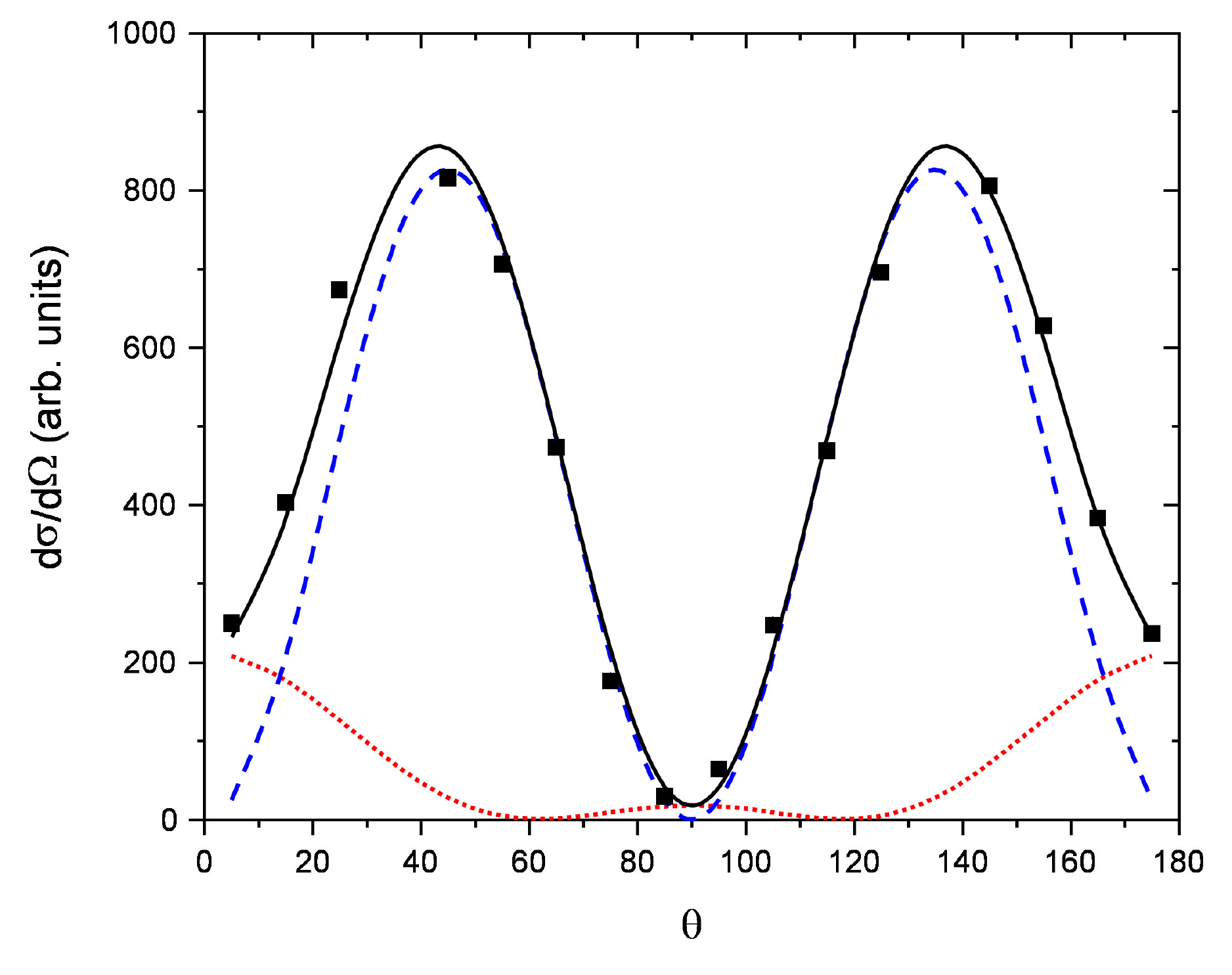}
\caption{Angular differential cross section d$\sigma$/d$\Omega$ as a function of the polar angle $\theta$ for a photon polarization in the $z$-direction. Full squares are experimental data, the dashed line is the calculation for an initial state with $m_l=\pm 1$, the dotted line is the calculation for $m_l=0$ multiplied by 0.05. The solid line represents the incoherent sum of both.\label{fig:ztheta}}
\end{figure}

\begin{figure}
\centering
\includegraphics[width=0.9\linewidth]{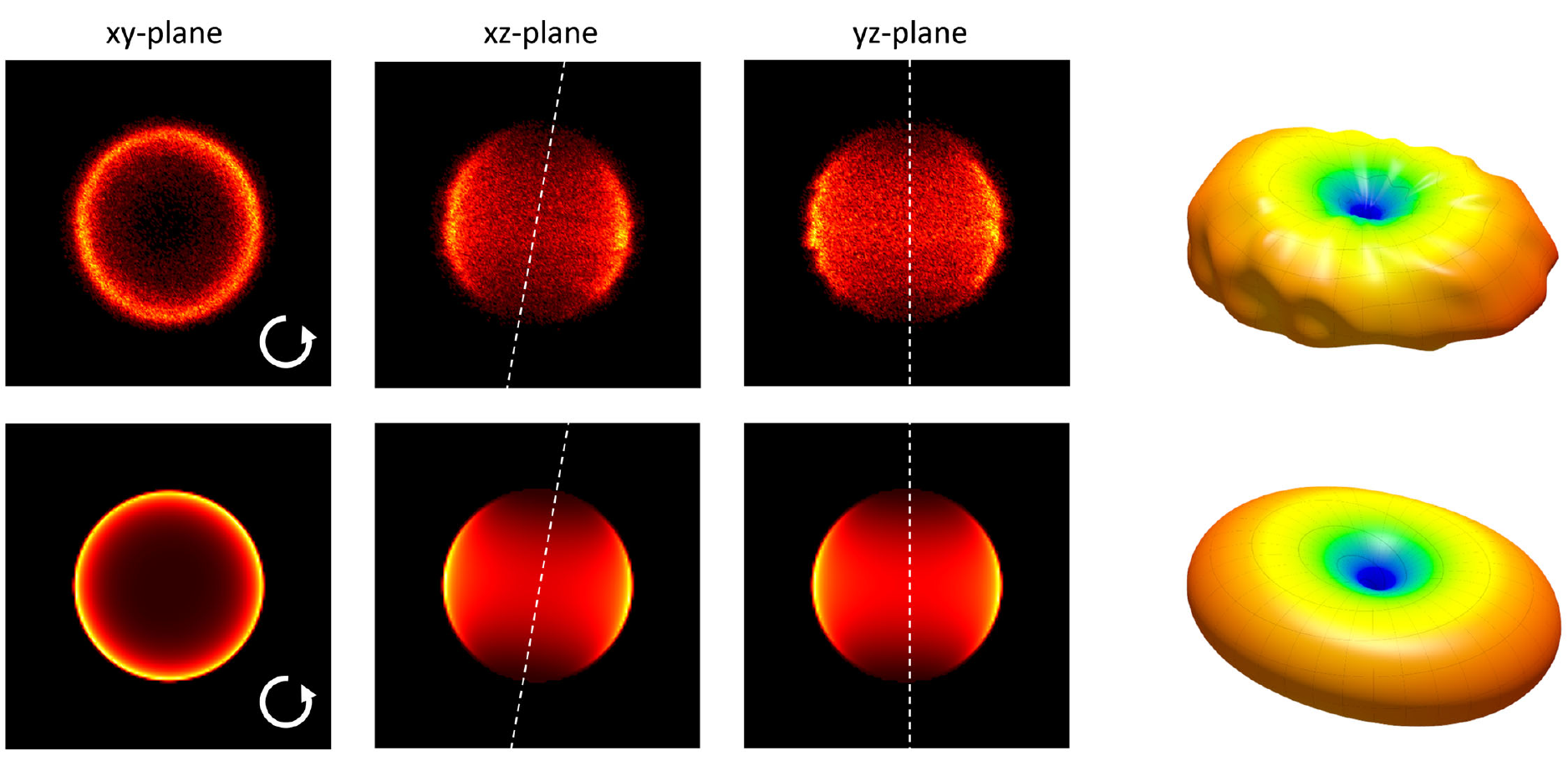}
\caption{Same as figure~\ref{fig:z} but for right-handed circular photon polarization in the $xy$-plane.\label{fig:rhc}}
\end{figure}

\begin{figure}
\centering
\includegraphics[width=0.5\linewidth]{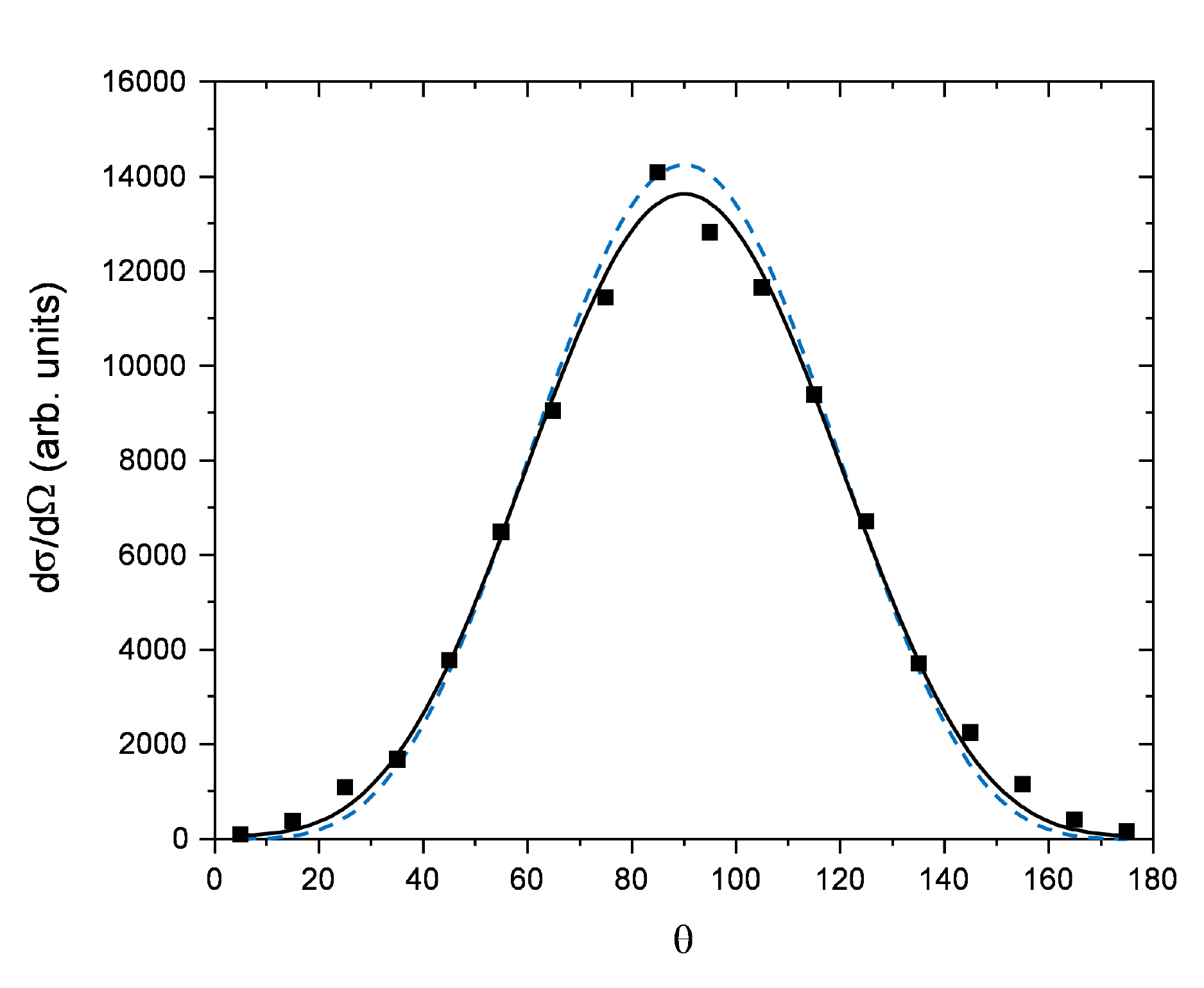}
\caption{Angular differential cross section d$\sigma$/d$\Omega$ as a function of the polar angle $\theta$ for a right-handed circular photon polarization in the $xy$-plane. Full squares are experimental data, the dashed line is the calculation for 100\% $m_l=+ 1$ initial state population. The solid line represents an incoherent sum with 93\%, 5\%, and 2\% of the atoms being initially in the $m_l=+1$, 0, and -1 magnetic sub-levels, respectively.\label{fig:rhctheta}}
\end{figure}

Generally, the deviation of the laser polarization from the $z$-direction results in a {\it coherent} admixture of partial waves of different $l$ and $m$ which induces essentially a slight rotation and a removal of strict cylindrical symmetry of the PAD. However, there can be {\it incoherent} contributions of different partial waves, too, caused by the incomplete polarization of the initial target state. In order estimate the degree of target polarization quantitatively, the cross section as a function of the electron momentum polar angle $\theta$ is shown in figure~\ref{fig:ztheta}, with the dashed line representing the calculation for 100\%  ($m_l=1$)-population of the target initial state. While there is excellent agreement between measured data and calculation for intermediate $\theta$, there are significant differences for angles smaller than about 40$^\circ$ and larger than 140$^\circ$. The calculation yields zero for electron emission exactly in positive or negative $z$-direction (i.e.\ $\theta=0^\circ$ and $180^\circ$, respectively),  which is not observed in the experiment. In an earlier study of the atom trap \cite{Sharma18}, an initial state $m_l$-distribution of roughly 0.93, 0.05, and 0.02 for $m_l=+1$, 0, and -1, respectively, was reported. Incoherently adding the differential cross section for the ionization from the 2p ($m_l=0$) state with a weight of 5\% (dotted line in the figure),  results in a nearly perfect match between calculation (solid line) and experimental data which is consistent with the earlier reported population ratio. The initial state with $m_l=-1$ does not need to be considered here, because it yields the same PAD as $m_l=+1$ for the present laser polarization.

For right-handed circular polarization and the laser beam propagating in the $z$-direction (i.e.\ $\Delta m=+1$), the calculation gets similarly simple as for the case above. Now, the only contributing final partial wave corresponds to the spherical harmonics $Y_{2}^{2}$. Experimental data and calculation for this polarization are shown in figure~\ref{fig:rhc}. The 10$^\circ$ angle of the laser beam with respect to the $z$-axis is accounted for by adding 17\% ($= \sin 10^\circ$) of electric field contribution to the $\epsilon_z$ component. The best agreement was achieved by additionally assuming a slight ellipticity of the radiation with an $\epsilon_-$ admixture of 6.5\%. The cross section as a function of the electron emission angle $\theta$ is shown in figure~\ref{fig:rhctheta}. Here the shape of the cross section is much less sensitive to incoherent contaminations due to the target's incomplete polarization, because such contribution are suppressed by their lower total cross sections for this specific laser beam polarization.

\begin{figure}
\centering
\includegraphics[width=0.9\linewidth]{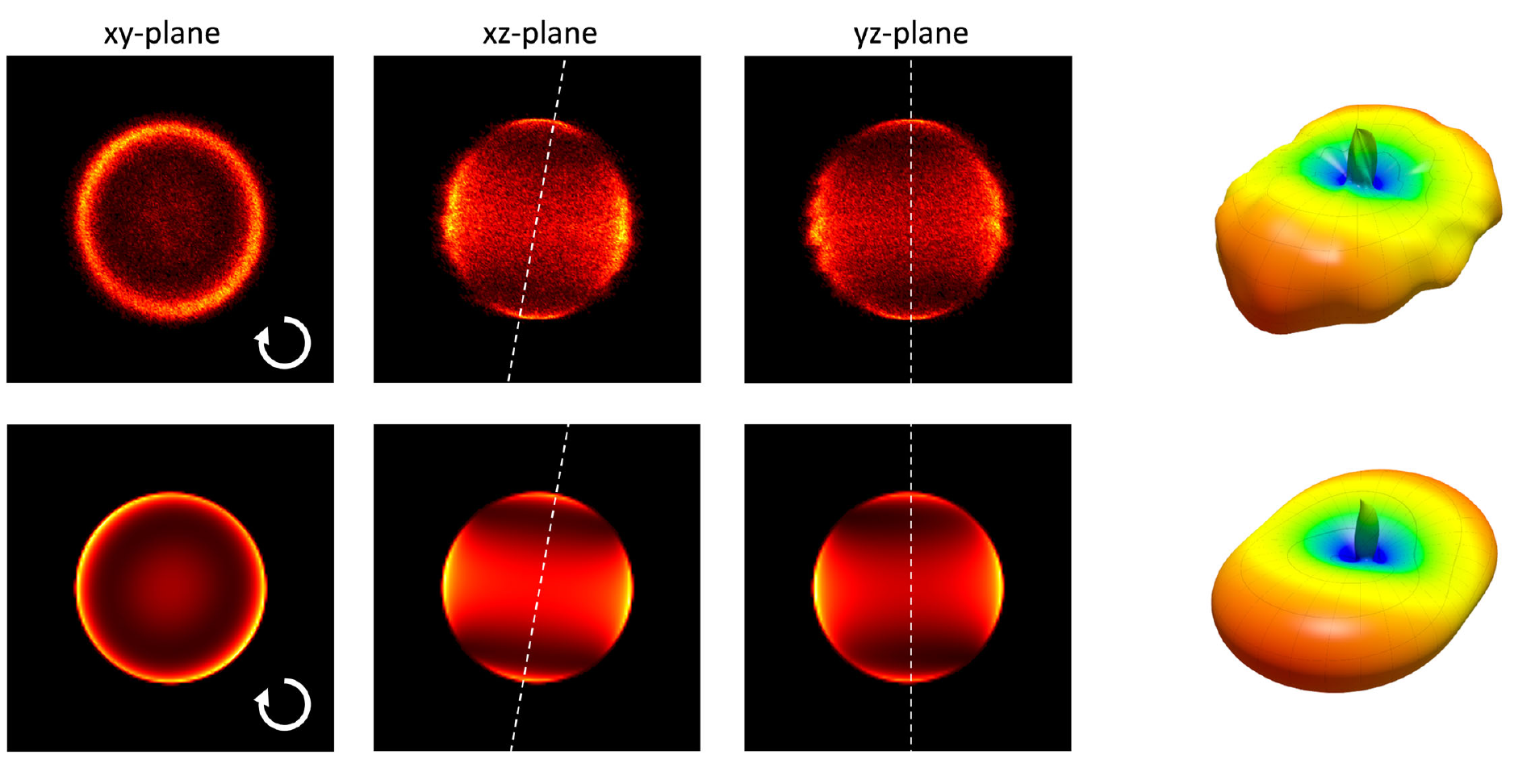}
\caption{Same as figure~\ref{fig:z} but for predominant left-handed circular photon polarization in the $xy$-plane.\label{fig:lhc}}
\end{figure}

The results for left-handed circularly polarized light moving along the $z$-axis is shown in figure~\ref{fig:lhc}. Here the dominant electric field component is $\epsilon_-$ resulting in the selection rule $\Delta m=-1$. Consequently, the final state is a superposition of two partial waves corresponding to $l=0$ and $2$ both with $m=0$. The relative amplitude and phase of the two partial waves was calculated in a one-electron model using the potential for lithium from \cite{Marinescu1994}. Similar to the case of right-handed circular polarization discussed above, the best match between experiment and calculation is achieved by including contributions of 17\% and 6.5\% for the $\epsilon_z$ and $\epsilon_+$ electric field components accounting for the beam angle and polarization ellipticity, respectively.

\begin{figure}
\centering
\includegraphics[width=0.5\linewidth]{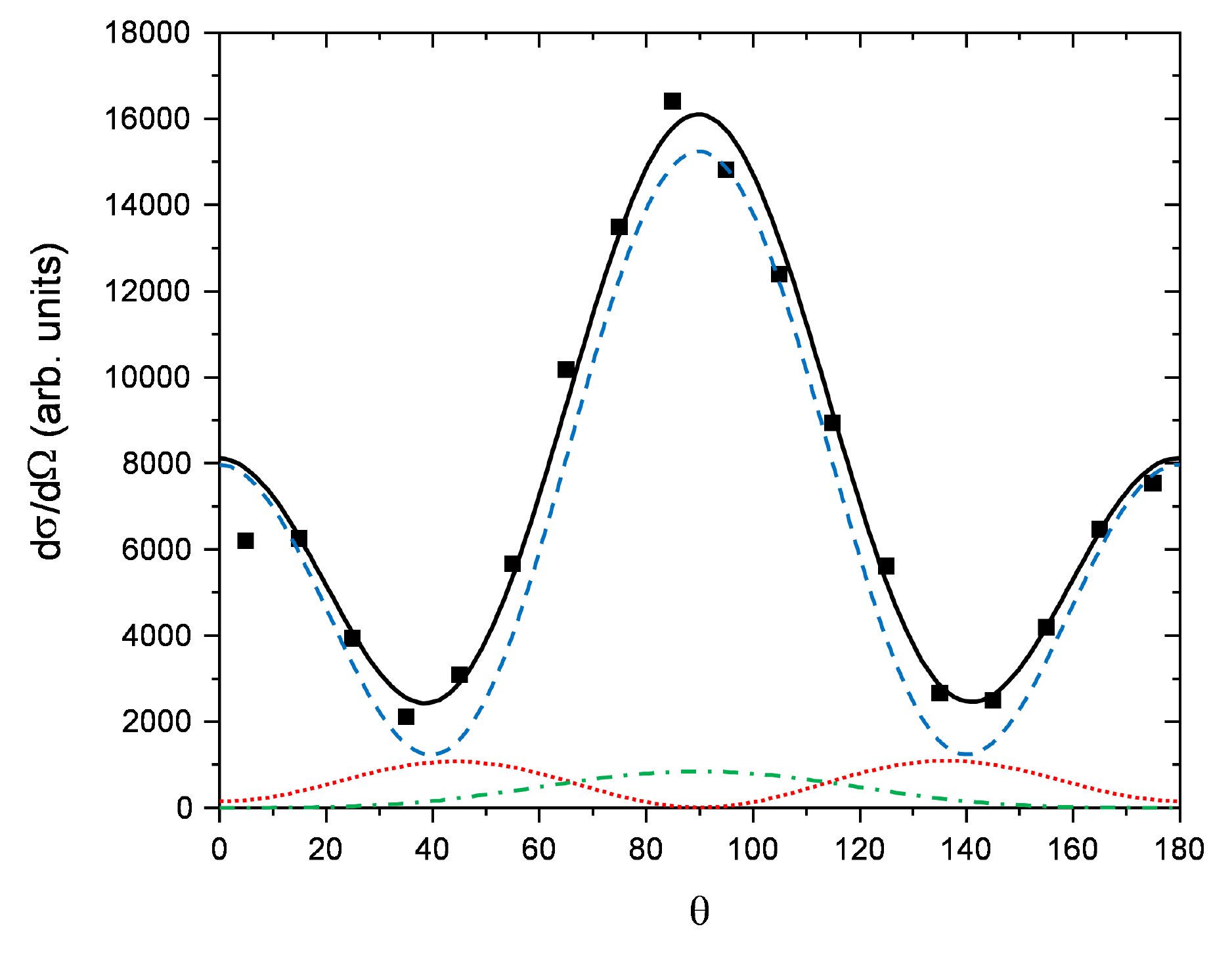}
\caption{Similar to figure~\ref{fig:rhctheta} but for left-handed circular photon polarization in the $xy$-plane. The solid line is the calculation for 93\% $m_l=+ 1$, 5\% $m_l=0$, and 2\% $m_l=-1$ initial state population. The individual contributions of the initial state magnetic sub-levels are shown as dashed, dotted and dash-dotted lines, respectively.\label{fig:lhctheta}}
\end{figure}

The $\theta$-distribution shown in figure~\ref{fig:lhctheta} features essentially three maxima at 0$^\circ$, 90$^\circ$, and 180$^\circ$. In principle, this distribution is very sensitive to incoherent admixtures of different target initial state polarizations and would allow to determine incoherent contributions of $m_l=-1$. Unfortunately, it is similarly sensitive to small variations of the relative amplitudes and phases of the involved partial waves. Because those parameters were calculated in a rather simple single-electron approximation, we hesitate to draw quantitative conclusions on the initial state polarization purity. Nevertheless, for the partial wave weighing factors used here, the comparison of the calculated and measured angular distributions appears to be consistent with the initial state $m_l$-population distribution reported earlier. 

\begin{figure}
\centering
\includegraphics[width=0.9\linewidth]{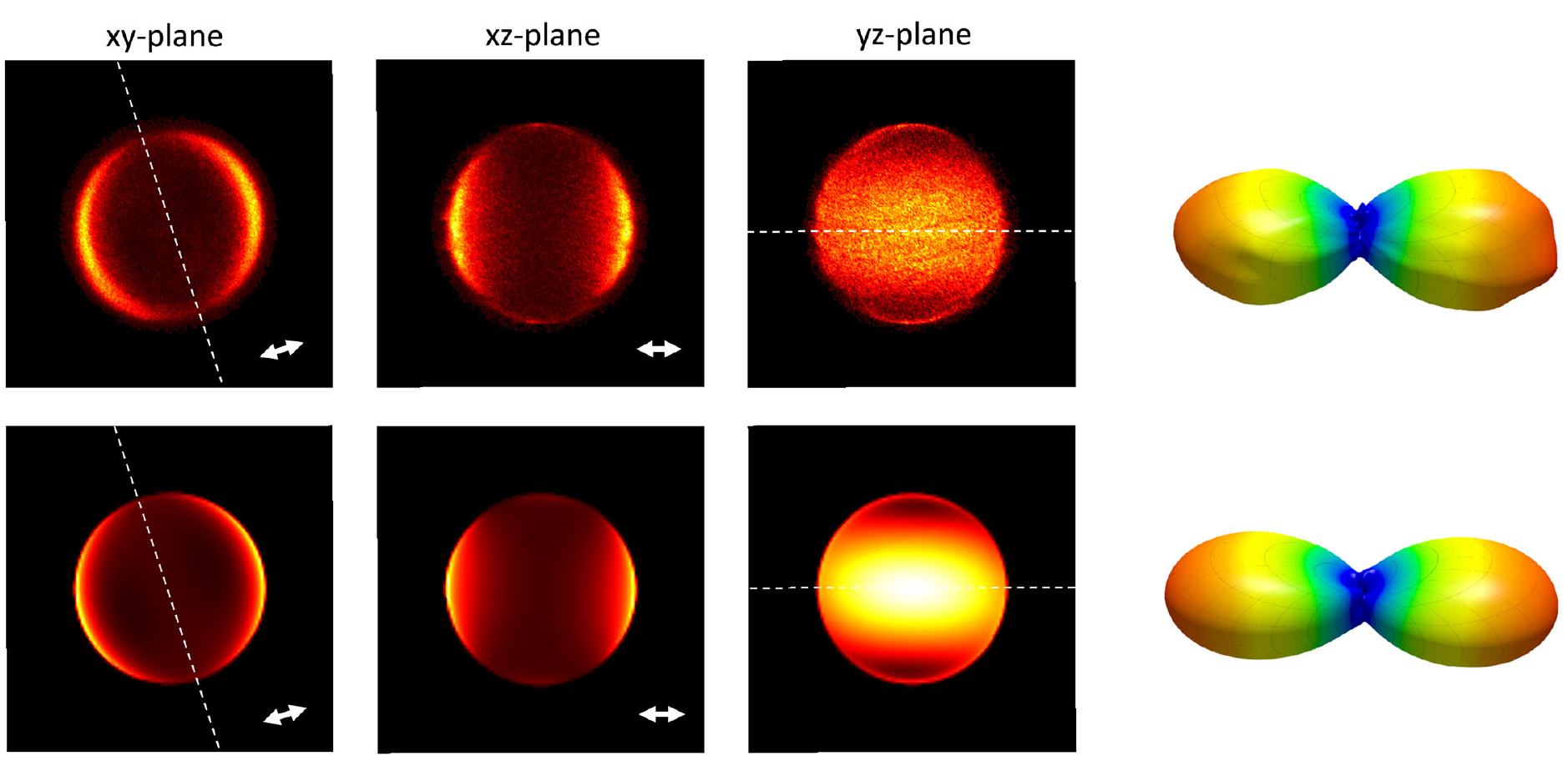}
\caption{Same as figure~\ref{fig:z} but for linear photon polarization perpendicular to the $z$-direction with an angle of 18$^\circ$ with respect to the $y$-axis.\label{fig:z90}}
\end{figure}

The fourth polarization under investigation is linear and perpendicular to the $z$-direction with an angle of 18$^\circ$ with respect to the $y$-axis. Here, $\epsilon_-$ and $\epsilon_+$ electric field components contribute equally and, consequently, the final electron wave function is a superposition of the spherical harmonics  $Y_{0}^{0}$,  $Y_{2}^{0}$, and  $Y_{2}^{2}$. The angular distribution of the emitted electrons features two pronounced peaks nearly aligned with the laser polarization direction as seen in figure~\ref{fig:z90}. 

\begin{figure}
\centering
\includegraphics[width=0.5\linewidth]{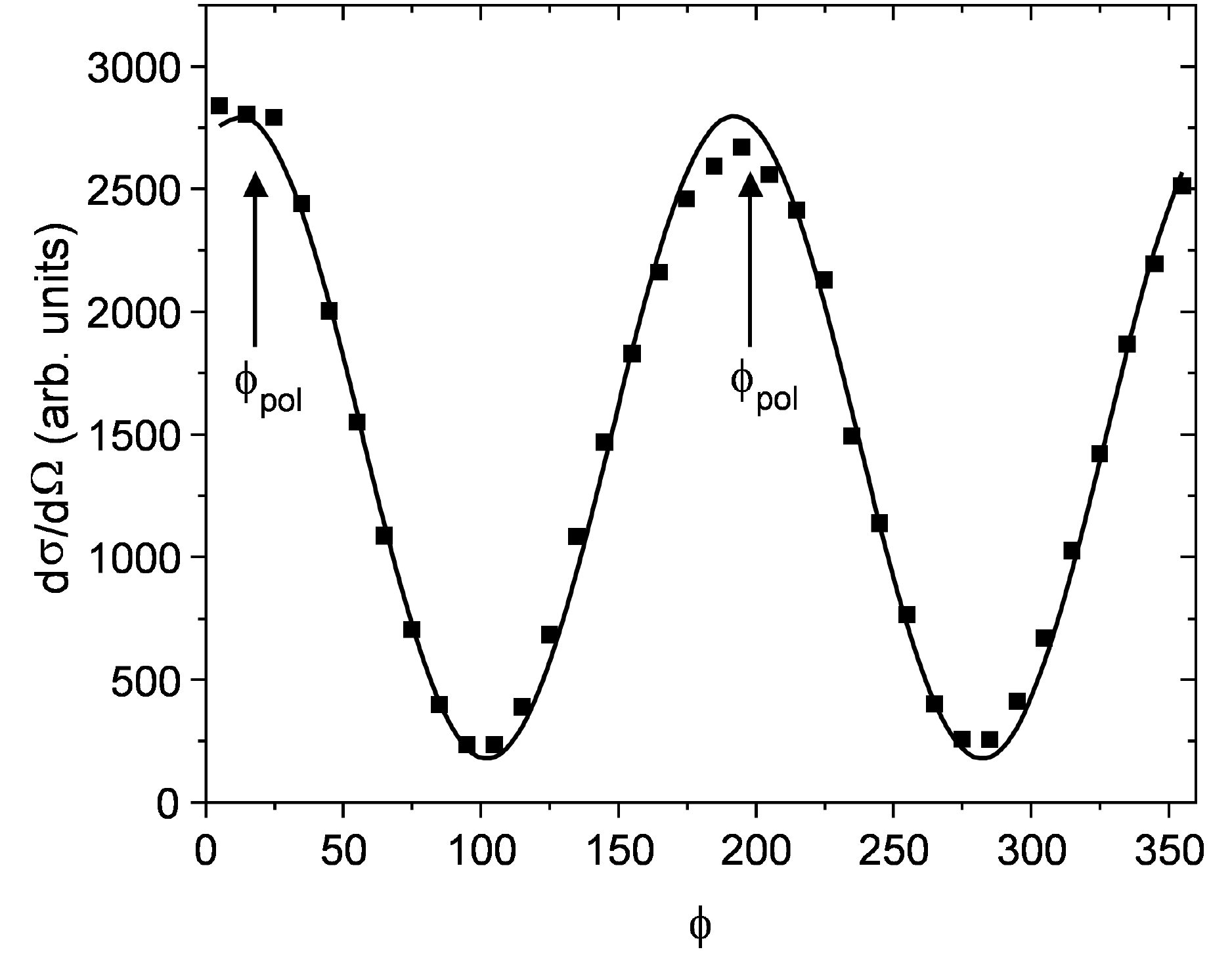}
\caption{PAD as a function of the azimuthal angle $\phi$ for electrons emitted in the $xy$-plane and for linear photon polarization perpendicular to the $z$-direction with an angle of 18$^\circ$ with respect to the $y$-axis. Squares are experimental data, the line is the fit according to Eq.~\ref{eq:z90}. The polarization direction is indicated by the arrows.\label{fig:z90phi}}
\end{figure}

It is worth noting that the PAD for this photon polarization is best suited to obtain information on the relative phase $\alpha_{02}$ of the partial waves with $l'=0$ and 2 from the measured data. In the above discussed case of left-handed circular photon polarization in the $xy$-plane, the amplitude and phase both affect the ratio of electron emission yield parallel and perpendicular the $z$-direction and there is no unambiguous fit that allows to extract the two parameters independently. For the present polarization, however, the strong $\phi$-dependence of the PAD allows to disentangle them. For electrons emitted in the $xy$-plane (i.e.\ $\theta=90^\circ$) the angular differential cross from Eq.~\ref{eq:PAD} becomes 
\begin{eqnarray}
&\frac{\rm{d}\sigma}{\rm{d}\Omega}(\theta =\pi/2, \phi)\propto& \left(5+2 A_{02}^2 +2 A_{02} \cos{\alpha_{02}} \right) + \nonumber\\
&& 3 \cos{2(\phi -\phi_{\rm pol})} + 6A_{02}\cos{(2(\phi -\phi_{\rm pol})-\alpha_{02})}
\label{eq:z90}
\end{eqnarray}
with $\phi_{\rm pol}$ being the azimuthal angle of the laser polarization.  As seen from this equation and figure~\ref{fig:z90phi}, the angular distribution follows an oscillation pattern with two peaks in opposite directions. The peak position relative to the polarization direction, although it is not entirely independent of $A_{02}$, depends mainly on the phase $\alpha_{02}$. The relative amplitude $A_{02}$ determines the minimum to maximum ratio of the $\phi$-distribution in figure~\ref{fig:z90phi}.

Although the deviation is very small in the present experiment, it is interesting to note that the peak in the electron emission pattern is generally shifted from the photon polarization direction (see Fig.~\ref{fig:z90phi}) if the phase shift $\alpha_{02}$ between the contributing final state partial waves does not vanish. This symmetry-breaking is not limited to photo-ionization but it has also been observed earlier for charged particle impact ionization of polarized alkali atoms. It appeared as 'orientational dichroism' in electron \cite{Dorn1998} and ion impact ionization \cite{Hubele2013} where the mean electron emission direction is shifted away from the momentum transfer direction. 

\begin{figure}
\centering
\includegraphics[width=0.9\linewidth]{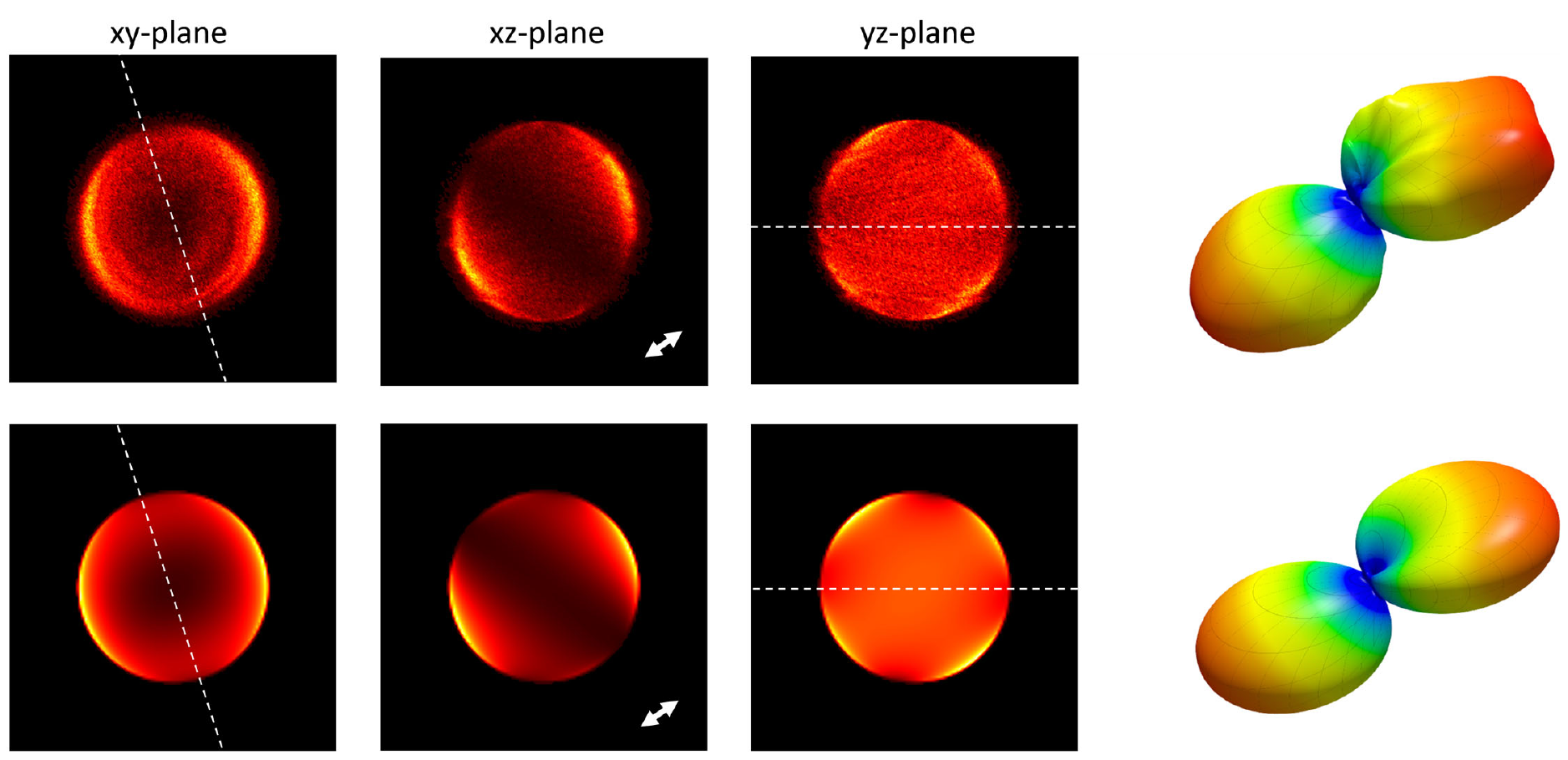}
\caption{Same as figure~\ref{fig:z} but for linear photon polarization with an angle of about 45$^\circ$ with respect to the $z$-axis (see text).\label{fig:z45}}
\end{figure}

Finally, a coherent superposition of all four possible partial waves was achieved by rotating the polarization by another 45$^\circ$  such that the polarization vector points in the (1/($\sqrt{2}$\,sin\,18$^\circ$), 1/($\sqrt{2}$\,cos\,18$^\circ$), 1/$\sqrt{2}$) direction. Also for this rather exotic case, a nearly perfect agreement between calculation and measurement is achieved (see Fig.~\ref{fig:z45}).

\section{Conclusion} 
In conclusion, we have measured photo-electron angular distributions in single photo-ionization of excited and polarized lithium atoms. The novel experimental approach consists of an all-optical atom trap in a COLTRIMS momentum spectrometer where the target atoms are laser-cooled and polarized, before they are momentum analyzed after ionization processes. In the present study, atoms in the excited 1s$^2$2p($m_l=1$) electronic configuration were ionized by the absorption of a single photon from a pulsed, low-intensity Nd:YAG laser source operating at a wavelength of 266\,nm. The analysis of the obtained photo-electron momentum distribution provides information not only on the details of the collision system but also on the characteristics of the experiment.     

Specifically, with the measured data the electron momentum resolution was determined to be 0.005\,a.u.\ (FWHM) in the $z$-direction and about 0.01 to 0.02\,a.u.\  perpendicular to it. Moreover, this excellent resolution allows to extract the relative phase and amplitude of the final state partial waves with angular momenta $l=0$ and 2. These two target species and photon energy specific parameters enable to reproduce consistently the photo-electron angular distributions for all investigated photon polarizations with excellent agreement. 

Minor deviations to the calculation were attributed to two systematic effects of the present experiment: First, there is a small birefringence of the fused-silica vacuum windows that slightly alters the polarization of the ionizing photons. To account for this effect, the polarizations used in the calculation were modified by a small margin, too, thereby optimizing the agreement between experiment and calculation. The influence of this effect is expected to be much stronger for the present ultra-violet radiation than for visible or infrared light, because the effective optical path through the glass per wavelength is unfavorably large for the ultra-violet photons, thereby resulting in larger phase shifts. Second, the target atoms are not fully polarized but there are small incoherent admixtures of magnetic sub-levels other than $m_l=1$. A 5\%-contribution of $m_l=0$ and 2\% of $m_l=-1$ were used in the calculation consistent with an earlier measurement of fluorescence imaging, and nearly perfect agreement to measured spectra were achieved.
 
The present study shows that the all-optical trap along with the momentum spectrometer allow to obtain high-resolution and high-quality data providing insights into detailed structures of the final momentum space. The setup allows to choose the laser propagation direction (and polarization) relative to the target, thereby controlling the angular momentum transfer, 
which can be used in future measurements where multi-photon or strong-field ionization of polarized lithium atoms, molecules, or ultra-cold atomic samples will be investigated.


\section*{Acknowledgments}
This material is based upon work supported by the National Science Foundation under Grant No.~1554776.

\section*{References}

\bibliographystyle{jphysicsB}

\end{document}